# Technological research in the EU is less efficient than the world average. EU research policy risks Europeans' future

Alonso Rodríguez-Navarro[a,b,*], Ricardo Brito[b]

*[a] Departamento de Biotecnología-Biología Vegetal, Universidad Politécnica de Madrid, Avenida Puerta de Hierro 2, 28040, Madrid, Spain*
*[b] Departamento de Estructura de la Materia, Física Térmica y Electrónica and GISC, Universidad Complutense de Madrid, Plaza de las Ciencias 3, 28040, Madrid, Spain*
*\* Corresponding author*
*e-mail addresses*: AR-N, alonso.rodriguez@upm.es; RB, brito@ucm.es

We have studied the efficiency of research in the EU by a percentile-based citation approach that analyzes the distribution of country papers among the world papers. Going up in the citation scale, the frequency of papers from efficient countries increases while the frequency from inefficient countries decreases. In the percentile-based approach, this trend, which is uniform at any citation level, is measured by the $e_p$ index that equals the $P_{top\ 1\%}/P_{top\ 10\%}$ ratio. By using the $e_p$ index we demonstrate that EU research on fast-evolving technological topics is less efficient than the world average and that the EU is far from being able to compete with the most advanced countries. The $e_p$ index also shows that the USA is well ahead of the EU in both fast- and slow-evolving technologies, which suggests that the advantage of the USA over the EU in innovation is due to low research efficiency in the EU. In accord with some previous studies, our results show that the European Commission's ongoing claims about the excellence of EU research are based on a wrong diagnosis. The EU must focus its research policy on the improvement of its inefficient research. Otherwise, the future of Europeans is at risk.

*Key words: research evaluation; percentile distribution; research efficiency; European research; citation; $e_p$ index.*

## 1. Introduction

Production and cost analysis plays a central role in the management of all productive systems, because it is the starting point for obtaining better and more profitable products. Research is also a productive process, for which production and costs should likewise be analyzed in order to improve its societal benefits. However, there are multiple examples of countries' research policies that are established without any production and cost analysis, either on the assumption that research is always profitable or taking for granted conclusions about its output that have never been demonstrated. The most remarkable case of the latter is the research policy of the EU.

For a long time, it has been held that the EU's technological weakness lies in its inferiority in transforming scientific breakthroughs and technological achievements into industrial and commercial successes; this has been known as the "European paradox" (European-Commission, 1995). The basis for this proposal is that EU's research is excellent and that the EU leads the world in scientific breakthroughs and technological achievements. This assumption of excellence has been the basis of EU research policy from 1995 through to the current EU research framework program Horizon 2020 but, surprisingly, research policy makers in the EU have never demonstrated the existence of such an excellence. In contrast with this political assumption, many academic publications have demonstrated that the proposed excellence of scientific research in the EU is highly questionable or in fact inexistent (Albarrán et al, 2010; Bonaccorsi, 2007; Bonaccorsi et al, 2017a; Dosi et al, 2006; Herranz & Ruiz-Castillo, 2013; Rodríguez-Navarro, 2016; Rodriguez-Navarro & Narin, 2017; Sachwald, 2015).

Currently, two main documents have been produced to give support to the new EU research program that will substitute for Horizon 2020: the "Interim Evaluation of Horizon 2020" (European-Commission, 2017a) and the "LAB-FAB-APP – Investing in the European future we want" (European-Commission, 2017b). In these documents, the "European paradox" is not mentioned, but the assumption of the excellence of the EU research is identical to which has always underpinned the idea of the "European paradox." Although it is well known that innovation goes beyond science and technology, and that incremental innovation might occur independently from basic research, for breakthrough innovation at the leading edge of knowledge, research is crucial (e.g., (OECD, 1996); (Leydesdorff, 2010). Therefore, it is highly worrying that the European Commission continues to apply a research policy that ignores academic findings, which indubitably demonstrate the weakness of EU research.

One factor that might explain the reluctance of the European Commission to accept the academic findings could be the complexity of academic approaches. To solve this problem a recently developed approach based on the well-established percentile apportionment method (Bornmann, 2013; Bornmann et al, 2013; Waltman & Schreiber, 2013) uses two simple indicators which are obtained by analyzing the distribution of country papers among the world papers (Brito & Rodríguez-Navarro, 2018). Going up in the percentile scale, the frequency of papers from the more research-active countries increases while the frequency from the less-active countries decreases. The trend of this frequency is uniform at any citation level and is measured by the first indicator used in this study, the $e_p$ index (see section 3). The second indicator, $P'_{top\ 0.01\%}$, estimates the likelihood for a research system to publish very highly cited papers (Brito & Rodríguez-Navarro, 2018). Although the $P'_{top\ 0.01\%}$ indicator estimates the frequency of infrequent events, it is calculated attending to the distribution of all publications, which includes the lowly cited ones that are the most numerous in all research systems.

By using these two new mathematically based indicators, this study aimed to answer the question of whether research in the EU is excellent, as proposed by the European Commission, or weak, as proposed by several academic publications. Furthermore, we centered this study on technology, performing bibliometric searches on the research topics that support technological advancements in the forefront of knowledge.

## 2. Metrics of research excellence

The question addressed in this study is whether the research excellence of EU research that is assumed by policy makers is actually true. Since 1995, when the existence of a "European paradox" was proposed (European-Commission, 1995), a large number of documents from the European Commission have praised the excellence of EU research. This praise continues in two current documents that are important for future EU research policy (European-Commission, 2017a; European-Commission, 2017b). In these documents, the number of sentences or paragraphs referring to research excellence that could be recorded is very large. However, this continuous application of the term *excellence* to research takes place without reference to any definition or metric. This absence of precision seems to be a general problem: the OECD document "Promoting Research Excellence. New approaches to funding" explicitly states: "The issue of what research excellence actually is or should be about is not part of this report" (OECD, 2014), p. 21). For more information the OECD document refers to the 2012 conference "Excellence Revised" (www.excellence2012.dk, accessed 01/10/2018) where a definition cannot be found.

In scientometrics, references to excellence are very frequent (e.g., (Bonaccorsi et al, 2017b; Tijssen et al, 2002); in 2014 there were more than 70,000 references to "research excellence" in research literature (Sorensen et al, 2016). However, in most cases, excellence is associated to a fuzzy concept for which "no single indicator of excellence can be used in isolation to capture the full picture" (Tijssen, 2003), p. 95).

Consistent with this fuzzy concept, a publication from the EU's Joint Research Center entitled "Composite Indicators or Research Excellence" (Vertesy & Tarantola, 2012) reports an exhaustive analysis of 22 indicators. More recently the European Commission has reduced the number of indicators to just four (Hardeman et al, 2013), of which only the first—number of highly cited publications—is bibliometric. The usefulness of the three non-bibliometric indicators: (i) high quality patent applications, (ii) world class universities and research institutions, and (iii) high prestige research grants, is not clear because they are related to or dependent on the bibliometric one. Thus, the relation of high quality patents and highly cited papers has been demonstrated by (Narin et al, 1997); it is probable that all universities and institutions research rankings take the number of highly cited publications into account; it is unlikely that high prestige research grants do not take into account the research experience of the project's authors, which correlates with the number of highly cited papers they have published. Taking these considerations together, the number of highly cited papers seems to be the most important indicator of research excellence.

Moreover, the use of the number of highly cited publications as an indicator of research excellence underlays the assumption that research excellence equates to a high scientific impact. However, although this impact may be estimated from the number of highly cited papers (Brito & Rodríguez-Navarro, 2018) and references therein), excellence implies superiority but does not indicate the magnitude of this superiority. Therefore, in most evaluations, the selection of the citation level or percentile threshold is made arbitrarily (Schreiber, 2013). In percentile-based evaluations, thresholds of 10%, 5%, or 1% have been widely used (Bonaccorsi et al, 2017a; Bornmann et al, 2015; Dosi et al, 2006; King, 2004; Leydesdorff et al, 2014; Tijssen et al, 2002; van-Leeuwen et al, 2003). In rankings, SCIMAGO (http://www.scimagoir.com/, accessed 01/12/2018) and "Mapping Scientific Excellence" (http://www.excellencemapping.net/, accessed in 01/12/2018) use the 10% threshold while the "Leiden Ranking" (http://www.leidenranking.com/, accessed 01/12/2018) provides the rankings for 50%, 10%, and 1% thresholds. The arbitrariness of percentile selection creates a problem because superiority can change into inferiority depending on the percentile selected (Fig. 2 in (Brito & Rodríguez-Navarro, 2018).

In the aforementioned European Commission documents, the 10% percentile is selected to measure research excellence (Hardeman et al, 2013), p. 18), but this selection is intrinsically misleading. In fact, in recent years the EU's concept of excellence has varied and now "excellence is more sharply defined and connected with a particular sort of knowledge that produces breakthroughs" (Sorensen et al, 2016), p. 217); it is highly questionable that 10 out of 100 publications report scientific breakthroughs.

In addition to the complex issue of the citation or percentile threshold from which excellence should be established, another issue is that although, in principle, excellence is a size-independent concept, its evaluation by some methods is size dependent (Crespo et al, 2012). In percentile-based assessments, a way to obtain an indicator independent from size is to divide the percentile counts by the total number of publications (PP$_{\text{top x\%}}$ indicators; (Bornmann et al, 2014; Waltman et al, 2012; Waltman & Schreiber, 2013)). This approach normalizes for the size, but if we compare two countries the ratio between the resulting indicators will vary depending on the percentile selected.

This variation can easily be demonstrated because the distribution of publications in percentiles follows a simple power law function that fits a wide range of percentiles (Brito & Rodríguez-Navarro, 2018). Then, the number of papers in a percentile $x$, denoted by $N(x)$, can be written as:

$$N(x) = A\, x^{\alpha} \qquad [1]$$

where $x$ ranges between 0% and 100%. In the percentile 100% all $N$ papers are included, so we can calculate $A$ and rewrite the equation above as:

$$N(x) = N \left(\frac{x}{100}\right)^{\alpha} \qquad [2]$$

Finally, dividing by $N$ we obtain the cumulative probability function that remains a function of $x$, but independent of N. Thus, when comparing the PP$_{\text{top x\%}}$ indicators of two countries, the result will depend on the percentile. For example, if we calculate the ratio between the PP$_{\text{top x\%}}$ indicators of two countries with $\alpha 1$ and $\alpha 2$ power law exponents, we have

$$\text{PP}_{\text{top x\%}} \text{ ratio} = \left(\frac{x}{100}\right)^{\alpha 1 - \alpha 2} \qquad [3]$$

this equation demonstrates that comparisons at different percentiles produce different results.

## 3. The $e_p$ index and other research excellence indicators

To solve this conundrum about the meaning of research excellence and the difficulties of its quantification, we propose to associate excellence with efficiency, which does have a quantitative meaning. However, this association implies neither a link with productivity—as in "Productivity is the quintessential indicator of efficiency in any production system" ((Abramo & D'Angelo, 2014), p. 1129)—nor any input-output relationship. This relationship has been extensively discussed elsewhere (Abramo & D'Angelo, 2016; Glanzel et al, 2016; Ruiz-Castillo, 2016; Waltman et al, 2016) but here we use efficiency in the sense of *intrinsic efficiency* or *breakthrough potential*, i.e., independent of inputs. This definition could be seen as the capacity of a research system to produce *revolutionary science* with the minimum possible amount of *normal science*, using Kuhn's terms (Kuhn, 1970). To make this definition of efficiency quantifiable we take advantage of the power law function that describes the percentile distribution of publications, equation [1].

Also important for the purpose of our study is that our definition of excellence in terms of *intrinsic efficiency* or *breakthrough potential* is not related to some definitions that have been used elsewhere. For example, "a national research system's efficiency can be defined as the extent to which a country is able to transform research assets into excellent research" ((Hardeman & van-Roy, 2013), p. 1). In this case, assets refer to gross expenditures in R&D, Government and higher education sector expenditures in R&D, and business expenditures in R&D, and excellent research is measured by the top 10% most cited papers, as previously explained.

To measure the *intrinsic efficiency* or *breakthrough potential* of a research system, here we describe the $e_p$ (**e**fficiency based on **p**ercentiles or **e**xcellence based on **p**ercentiles) index. The $e_p$ index is defined as

$$e_p = P_{top\ 1\%}/P_{top\ 10\%} \qquad [4]$$

operating with the values of the $P_{top\ 1\%}$ and $P_{top\ 10\%}$ indicators shown in equation [2] we obtain

$$e_p = 10^{-\alpha} \qquad [5]$$

where $\alpha$ is the exponent in equations [1] and [2]. At any percentile level, this exponent determines the potential of a country or institution in producing highly cited publications—higher $\alpha$ implies lower efficiency because $x/100$ is ≤ 1. In other words, because highly cited papers cannot exist without a quite high amount of lowly cited papers, the $e_p$ index measures the *intrinsic efficiency* of the research system in producing papers at a given citation level from the number of papers at lower levels of citations; for example, let us say, the number of the top 10% most cited papers with reference to the number of the top 50% most cited papers. For a research system of any size that is identical to the world system—identical $\mu$ and $\sigma$ parameters of the lognormal distributions (Rodríguez-Navarro & Brito, 2018)—the value of the $\alpha$ exponent is 1.0; consequently the value of the $e_p$ index is 0.10 (equation [5]).

The $e_p$ index measures the *intrinsic efficiency* or *breakthrough potential* in comparative terms, because it measures the distribution of the papers from a particular country across different layers of citations with reference to the world papers. Thus, the $e_p$ index implies success in a competition, in coincidence with the concept of excellence, which implies to be *superior* in its class.

The proportion the world's discoveries or breakthroughs achieved by a country depends on two terms, the size and *breakthrough potential* of its research system or, in other words, on its size and value of the $e_p$ index. Therefore, to have a strong research system, countries should try to achieve the highest possible $e_p$ index and to increase the size of their system, but never to increase the size ignoring the $e_p$ index.

In addition to the $e_p$ index, our results also record the $P_{top\ 0.01\%}$ indicator. A $P_{top\ x\%}$ indicator is useful for research evaluation because it indicates the capacity of the system to publish papers with a certain citation level, which implies the production of breakthroughs of certain relevance. Then, the indicator may be normalized with reference to other parameters such as population size, volume of research investments, or GDP. The selection of the $P_{top\ x\%}$ indicator for evaluation purposes is not necessarily arbitrary (Schreiber, 2013). The *x%* level depends on the opinion of experts within a field about the annual number of breakthroughs, paradigm shifts, or important discoveries that can be expected. The ratio between the number of these achievements and the total number of papers determines the percentile to be used. For example, with 100,000 publications in a field, the use of the 1% percentile would imply 1,000 such achievements; our use of the 0.01% assumes 10. We used the $P_{top\ 0.01\%}$ indicator, in the first place, because in the field of chemistry it correlates with the number of achievements in the USA and the EU awarded with Nobel Prizes (Brito & Rodríguez-Navarro, 2018), and, in the second place, because it implies a reasonably number of important breakthroughs, as we describe in the next section.

Given the considerations described above in this section, the $P_{top\ x\%}$ indicators can be obtained in two different ways, either by counting the papers or by using equation [1] after fitting empirical data points to this equation. Therefore, we hereafter use the notations $P_{top\ x\%}$ and $P'_{top\ x\%}$ to distinguish indicators that have been obtained by counting and calculation, respectively. However, we use a single notation for the $e_p$ index regardless of whether it was calculated from the $P_{top\ 10\%}$ and $P_{top\ 1\%}$ indicators or by curve fitting.

## 4. Preliminary data and study design

Table 1. Research evaluation based on the $e_p$ index, P'$_{top\,0.01\%}$ indicator, and P'$_{top\,0.01\%}$ per million inhabitants across countries[a]

| Country | P$_{top\,10\%}$ | P$_{top\,1\%}$ | $e_p$ index | P'$_{top\,0.01\%}$ | P'$_{top\,0.01\%}$ per million inhabitants |
|---|---|---|---|---|---|
| Switzerland | 49,275 | 5,859 | 0.119 | 82.84 | 9.86 |
| Denmark | 25,022 | 2,832 | 0.113 | 36.28 | 6.25 |
| US | 858,703 | 96,146 | 0.112 | 1205.33 | 3.71 |
| The Netherlands | 64,667 | 7,060 | 0.109 | 84.15 | 4.92 |
| Austria | 17,785 | 1,919 | 0.108 | 22.34 | 2.54 |
| Belgium | 29,419 | 3,102 | 0.105 | 34.49 | 3.03 |
| Canada | 100,307 | 10,474 | 0.104 | 114.20 | 3.13 |
| Norway | 14,312 | 1,493 | 0.104 | 16.25 | 3.07 |
| Sweden | 41,792 | 4,327 | 0.104 | 46.38 | 4.64 |
| UK | 201,588 | 20,855 | 0.103 | 223.20 | 3.39 |
| Finland | 18,247 | 1,837 | 0.101 | 18.62 | 3.39 |
| Australia | 58,612 | 5,854 | 0.100 | 58.40 | 2.40 |
| New Zealand | 10,361 | 1,026 | 0.099 | 10.06 | 2.14 |
| Germany | 159,250 | 15,738 | 0.099 | 153.71 | 1.86 |
| Israel | 22,266 | 2,180 | 0.098 | 20.90 | 2.43 |
| Poland | 12,042 | 1,170 | 0.097 | 11.04 | 0.29 |
| France | 112,965 | 10,971 | 0.097 | 103.48 | 1.60 |
| Italy | 74,378 | 7,150 | 0.096 | 66.07 | 1.09 |
| South Africa | 7,159 | 661 | 0.092 | 5.64 | 0.13 |
| China | 75,537 | 6,827 | 0.090 | 55.77 | 0.04 |
| Greece | 10,134 | 913 | 0.090 | 7.41 | 0.64 |
| Spain | 50,797 | 4,526 | 0.089 | 35.93 | 0.77 |
| Russia | 15,887 | 1,413 | 0.089 | 11.18 | 0.08 |
| Mexico | 6,169 | 531 | 0.086 | 3.93 | 0.04 |
| Japan | 109,249 | 9,371 | 0.086 | 68.95 | 0.54 |
| Brazil | 16,025 | 1,309 | 0.082 | 8.73 | 0.05 |
| Korea | 25,233 | 2,037 | 0.081 | 13.28 | 0.27 |
| Turkey | 10,100 | 793 | 0.079 | 4.89 | 0.07 |
| Taiwan | 18,612 | 1,332 | 0.072 | 6.82 | 0.30 |
| India | 22,320 | 1,530 | 0.069 | 7.19 | 0.01 |

[a] The P$_{top\,10\%}$ and P$_{top\,1\%}$ indicators were taken from (Bornmann et al., 2015) and the P$_{top\,0.01\%}$ from (Brito and Rodríguez Navarro, 2018). The $e_p$ index, which reveals excellence, is equal to the P$_{top\,1\%}$/ P$_{top\,10\%}$ ratio (see text, section 3).

In several publications and rankings the P$_{top\,1\%}$ and P$_{top\,10\%}$ indicators have been reported (e.g., the Leiden Ranking). In these cases, the parameters of equation [1] can be easily obtained (Brito & Rodríguez-Navarro, 2018), $A = P_{top\ 1\%}$ and $\alpha = \lg(P_{top\ 10\%}/ P_{top\ 1\%})$, which implies that the $e_p$ index and the $P_{top\ 0.01\%}$ indicator can also be easily calculated.

To obtain a preliminary assessment of the level of research in the EU from published data, we calculated the $P_{top\ 0.01\%}$ indicator and $e_p$ index from the $P_{top\ 10\%}$ and $P_{top\ 1\%}$ data reported by Bornmann et al. (2015); we also added another column showing the $P'_{top\ 0.01\%}$ divided by number of inhabitants of the country (Table 1). The results show very little differences in the $e_p$ index across countries but higher differences the $P'_{top\ 0.01\%}$ indicator per million inhabitants. At the top of the rankings is Switzerland on both the $e_p$ index and the $P'_{top\ 0.01\%}$ per million inhabitants. Several other countries in the European Research Area (ERA): Denmark, The Netherlands, Sweden, Belgium, Norway, and Austria show higher indicators than the four biggest EU countries: Germany, France, Italy, and Spain.

Table 2. The $P_{top\ 0.01\%}$ indicator and $e_p$ index of leading universities across countries[a, b]

| University | Country | $P_{top\ 10\%}$ | $P_{top\ 1\%}$ | $e_p$ index | $P_{top\ 0.01\%}$ |
|---|---|---|---|---|---|
| Stanford University | USA | 976 | 169 | 0.173 | 5.06 |
| Massachusetts Institute of Technology | USA | 1175 | 184 | 0.156 | 4.50 |
| University of Cambridge | UK | 801 | 114 | 0.142 | 2.29 |
| Ecole Polytech Federale de Lausanne | Switzerland | 581 | 80 | 0.138 | 1.53 |
| ETH Zurich | Switzerland | 678 | 80 | 0.118 | 1.13 |
| Delft University of Technology | Netherlands | 395 | 56 | 0.141 | 1.11 |
| Imperial College London | UK | 598 | 70 | 0.117 | 0.95 |
| Technical University of Denmark | Denmark | 382 | 47 | 0.123 | 0.70 |
| University Paris XI-Paris-Sud | France | 314 | 38 | 0.121 | 0.56 |
| University Paris VI-Pierre and Marie Curie | France | 333 | 38 | 0.115 | 0.50 |
| RWTH Aachen University | Germany | 349 | 39 | 0.111 | 0.48 |
| Karlsruhe Institute of Technology | Germany | 414 | 42 | 0.102 | 0.44 |
| Katholieke Universiteit Leuven | Belgium | 318 | 35 | 0.111 | 0.43 |
| Ghent University | Belgium | 270 | 31 | 0.113 | 0.39 |
| University of Padova | Italy | 184 | 23 | 0.126 | 0.37 |
| University of Lisbon | Portugal | 254 | 27 | 0.108 | 0.32 |
| Universidad Autonoma de Madrid | Spain | 148 | 18 | 0.121 | 0.26 |
| Sapienza University of Rome | Italy | 205 | 22 | 0.106 | 0.24 |
| Universitat Politècnica de València | Spain | 159 | 18 | 0.112 | 0.23 |

[a] The $P_{top\ 10\%}$ and $P_{top\ 1\%}$ data were taken from the Leiden Ranking (http://www.leidenranking.com/ranking/2017/list) "Physical sciences and engineering." The universities selected were the first in each country from the Leiden Ranking ordered by the $P_{top\ 1\%}$ indicator. The $e_p$ index is equal to the $P_{top\ 1\%}/P_{top\ 10\%}$ ratio and means percentile-based excellence index as described in text, section 3.
[b] The $P_{top\ 0.01\%}$ indicator was calculated from $P_{top\ 10\%}$ and $P_{top\ 1\%}$ indicators according to the power law equation [1].

In advanced technological nations, universities play a central role in country's research (Godin & Gingras, 2000). Therefore, to continue our preliminary research assessment from published data, we calculated the $e_p$ index and the P'$_{top\ 0.01\%}$ indicator for leading universities in the USA and the EU, taking the P$_{top\ 10\%}$ and P$_{top\ 1\%}$ data reported by the Leiden Ranking (http://www.leidenranking.com/ranking/2017/list). Because our field of interest was technology, we performed our calculations on the data in the field of "Physical sciences and engineering" and selected the best universities in each country, ranked according to the P$_{top\ 1\%}$ indicator. Table 2 shows that according to both the $e_p$ index and the P'$_{top\ 0.01\%}$ indicator, the research performance of the two top universities in the USA was much better than that of the top universities in the ERA. In terms of the P'$_{top\ 0.01\%}$ indicator, the performance of the top universities in Germany, France, Italy, and Spain was poor in comparison with top USA universities, and, again, the universities of the UK, Switzerland, Denmark, and Netherlands were clearly ahead of other ERA (European Research Area) universities. In contrast with the data shown in Table 1, the values of the $e_p$ index in Table 2 are all higher than 0.1, which indicates that the selected universities have a better research performance than the world reference.

Although interesting, the results shown in Tables 1 and 2 do not reveal the defective research capacity of the EU at the leading edge of technological knowledge. For example, in Table 1, the P'$_{top\ 0.01\%}$ indicator should reveal the difference between the USA and the EU in garnering Nobel Prizes in Chemistry and Physics, but this does not occur because the USA/EU P'$_{top\ 0.01\%}$ indicator ratio is 1.3 (not shown calculations), which is much smaller than the expected ratio of 3.0 (Rodríguez-Navarro, 2016). In Table 2, a similar mathematical comparison is not possible, but again, according to previous analyses (Rodríguez-Navarro, 2012), the difference between the USA and Spanish or Italian universities should be higher than the 15-20 ratio that is obtained from the data in Table 2.

In both cases, these discrepancies occur because EU research is not homogeneously weak; it is competitive in areas of slow growth and uncompetitive in hot technological topics (Bonaccorsi, 2007; Rodriguez-Navarro & Narin, 2017; Sachwald, 2015), which makes the results very sensitive to the form in which the P$_{top\ 10\%}$ and P$_{top\ 1\%}$ papers are counted. When the publications in a broad research area, such as chemistry or physics, are analyzed as a totality, as in (Rodríguez-Navarro, 2016), hot topic publications are important determinants of the differences between USA and EU research because they are highly cited and more abundant in the USA. This higher abundance increases the $\mu$ parameter of the lognormal distribution of USA publications and, consequently, the USA-EU differences in the P'$_{top\ 0.01\%}$ indicator also increase. In contrast, when the analysis is performed independently in many narrow research areas, as in the Leiden Ranking or in (Bornmann et al, 2015), the final indicator is dominated by the performance in quiescent topics, which represent a high proportion of world's research, and in which the differences between the USA and EU research are low or inexistent.

These observations indicate that in order to obtain a reliable diagnosis of the EU research system our study has to analyze hot and quiescent areas independently without averaging the results. On the other hand, the most relevant studies should be performed in technological fields that are the most likely to produce the breakthroughs that support the knowledge-based economy. This notion is summarized with the sentence "The sort of research that matters is thus the kind that can deliver high financial returns through scientific breakthroughs and their commercialization" (Sorensen et al, 2016), p. 224). Therefore, in the first place, we identified research topics at the forefront of technological research. For this purpose, in a preliminary study we identified 14 fast evolving technological topics (hereafter referred to as FETT) from among a large number of research fields by their high citation rates and current technological importance. These topics were: graphene, solar cells, nanotechnology, electronics, $Li^+$ or $Na^+$ batteries, metal-organic frameworks, superconductors, transistors, semiconductors, wireless communications, composite materials, quantum dots, fuel cells, and energy transfer. Obviously, these 14 topics do not cover all possible FETT, but make up a large representative sample of research areas at the forefront of knowledge that supports current breakthrough innovations and will continue supporting it in the near future.

Furthermore, because it is known that EU research is weak in several of the selected FETT (Bonaccorsi, 2007; Rodriguez-Navarro & Narin, 2017; Sachwald, 2015), we next studied the EU research outputs in more traditional slow-evolving technological topics (hereafter referred to as SETT). For this purpose we selected 10 "research areas" in the broad WoS category of "technology": mechanics, engineering, materials science, energy & fuels, electrochemist, robotics, metallurgy & metallurgical engineering, automation & control systems, instruments & instrumentation, operation research & management science, and telecommunications. In the SETT searches we specifically exclude the research on FETT in order to produce two totally independent sets of topics. We also exclude "computer science," which we found requires a specific study. As explained for FETT, the selected SETT do not cover all possible slow-evolving technological topics, but collectively make up a large representative sample of them.

The worldwide number of articles in FETT and SETT were very similar, 194,147 and 180,196, respectively, in 2014. In contrast, citation distributions were very different. For example, in SETT articles, only five (0.003%) received more than 300 citations while in FETT articles, 207 (0.1%) exceeded that number of citations (searches on January 5, 2018). For this number of articles the use of the $P'_{top\ 0.01\%}$ indicator for research evaluation implies that in these fields there are approximately 18-19 important breakthroughs per year.

## 5. Methods

For the purpose of our study we divided the world into three geographical research areas: ERA, USA, and Others (i.e., all countries excluding ERA and USA). These areas were analyzed independently, omitting collaborative publications between them. In addition, in view of the large differences that exist in research between Switzerland or the UK and other ERA countries (Table 1), besides the searches for ERA countries, we also analyzed EU countries excluding the UK.

Bibliometric searches were performed in the Science Citation Index Expanded of the Web of Science Core Collection (WoS), using the "Advanced Search" feature and retrieving only publications labeled as "Articles" (this implies that review publications are not counted). For FETT we used: TS=(energy transfer OR fuel cell* OR quantum dot* OR composite material* OR transistor* OR semiconductor* OR superconductor* OR graphene OR batter* OR solar cell* OR electronic* OR nano* OR metal organic framework* OR wireless). For SETT we used: SU=((mechanics OR engineering OR materials science OR energy & fuels OR electrochemistry OR robotics OR metallurgy & metallurgical engineering OR automation & control systems OR instruments & instrumentation OR operations research & management science OR telecommunications) NOT computer science); we used the Boolean operator NOT to exclude FETT. We determined the number of publications in nine percentiles: 1, 2, 4, 7, 12, 20, 35, 60, and 100% by ordering publications according to their number of citations. The ordering of publications with the same number of citations was that provided by the database. To assign countries' publications at a specific world percentile we used the worldwide and country lists provided by the database and the number of citations to determine the country percentile limits. When the threshold occurred in a series of publications with the same number of citations both in the world and in the country, the country's threshold was situated by the proportional method as described previously (Brito & Rodríguez-Navarro, 2018).

For fractional counting, the downloaded publications from WoS were fractionally counted based on country's author affiliations. In a few cases, we found authors with affiliations in two countries and we counted these papers as fractional = 1 for both countries. However, we investigated this issue and found that it is irrelevant. Even in Singapore, where we found the highest number of double affiliations, counting a paper as either 1 or 0 had irrelevant effects for the purpose of this study.

All percentile-based citation distributions were analyzed by fitting to power laws as described previously (Brito & Rodríguez-Navarro, 2018). Deviations from a power law at the 100% and (rarely) 60% percentiles were observed for countries with very high or very low $e_p$ values; these data points were omitted for the fitting. In small and uncompetitive countries the number of papers in the 1–4% percentiles was very low and noisy and these data points were also omitted. The omission of these data points did not have any effect in our study because the fits of percentiles data points to power laws show very small deviations. This accuracy has been described previously and can be checked by the visual inspection of log-log plots (e.g., Fig. 1a and b). Furthermore, it is demonstrated by the high $R^2$ values of the fittings: we fit 64 power laws, of which 91.5%, 37.3%, and 8.5% produced $R^2$ values that were higher than 0.99, 0.999, and 0.98, respectively. In only one case, Poland, the log-log plot of the data points showed a biphasic trend that could not be fitted to a single power law. This biphasic trend can be analyzed in several ways but none of these analyses suggests that Poland has a significant role in EU research. Therefore, in section 6.1, Poland was omitted in the study of independent countries but is included in all other searchers as an EU or ERA country.

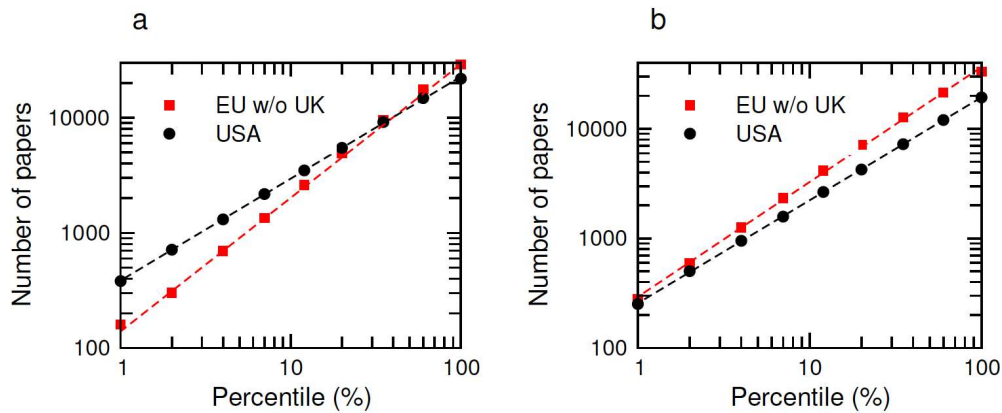

*Fig. 1. Percentile distribution of publications from the EU without the UK and the USA in 2014 in a) fast evolving technological topics and b) slow evolving technological topics.*

Although the $e_p$ index can be calculated in two ways, equations [4] and [5], we normally calculated it from the value of $\alpha$ that had been calculated by curve fitting.

Finally, for publications funded by the ERC (European Research Council) we used the tag FT=ERC.

## 6. Results

*6.1. Research performance in European countries*

To determine the research performance level in Europe, we calculated the $e_p$ index and the P'$_{top\ 0.01\%}$ indicator (section 3) for the four aforementioned geographical research areas for FETT and SETT publications in years 2012–2014. The results are recorded in Tables 3 and 4 for FETT and SETT publications, respectively. Both tables also record the total number of publications.

In FETT publications (Table 3), which represent the forefront of technological progress, the research performance of the EU without the UK was poor. The value of the P'$_{top\ 0.01\%}$ indicator, which reveals the likelihood of a country or institution publishing a highly cited paper, is approximately 10 times lower than in the USA. In comparison with the EU without the UK, in the ERA the P'$_{top\ 0.01\%}$ indicator increased two-fold, albeit the number of papers increased only 26%, which indicated than one or several countries in the ERA showed a much better research performance than the EU without the UK—these countries were obviously Switzerland and the UK (Table 1). In comparison with the Others, the EU's performance is again relatively poor: the findings are similar to those from the comparison with the USA because, according to the P'$_{top\ 0.01\%}$ indicator, the USA and Others are similar.

Table 3. Research performance based on the $e_p$ index and P'$_{top\ 0.01\%}$ indicator of four research world areas in fast evolving technology topics in three consecutive years[a]

| Research area | 2012 | | | 2013 | | | 2014 | | |
|---|---|---|---|---|---|---|---|---|---|
| | Papers | $e_p$ | P'$_{top\ 0.01\%}$ | Papers | $e_p$ | P'$_{top\ 0.01\%}$ | Papers | $e_p$ | P'$_{top\ 0.01\%}$ |
| USA | 20818 | 0.122 | 5.46 | 19394 | 0.125 | 6.00 | 19344 | 0.128 | 6.28 |
| ERA | 32949 | 0.074 | 1.08 | 42556 | 0.071 | 1.02 | 41889 | 0.080 | 1.58 |
| Others | 80568 | 0.090 | 4.99 | 85847 | 0.086 | 5.20 | 91834 | 0.087 | 6.13 |
| EU w/o UK | 26107 | 0.067 | 0.59 | 33500 | 0.062 | 0.48 | 33242 | 0.068 | 0.66 |

[a] The P'$_{top\ 0.01\%}$ indicator was calculated after fitting the percentile empirical data to a power law function. The $e_p$ index was calculated from de exponent of the power law equation, $e_p = 10^{-\alpha}$.

Using the $e_p$ index, the first finding was that the EU without the UK again showed a weak research performance. Remarkably, the $e_p$ index was lower in both the EU without the UK and the ERA than in Others. The second finding was that, although the P'$_{top\ 0.01\%}$ indicators in the USA and Others were similar, the $e_p$ index was much higher in the USA, which implies that for a similar P'$_{top\ 0.01\%}$ indicator the total number of publications was 4-5 times lower.

In SETT publications (Table 4), the scientific landscape was different. The first conclusion was that, in this case, significant differences between the EU without the UK and the ERA existed only in the number of papers. In both cases the $e_p$ index was not appreciably different from 0.10, which was lower than that of the USA (0.12) but higher than that of Others (0.09). While the $e_p$ index did not change appreciably over the three years of the study, 2012-2014, the P'$_{top\,0.01\%}$ indicator suffer important changes, which were due to the high increase in the number of publications from Others, 18.6% in three years.

Taken together the results revealed an alarmingly low level of technological research in the EU, especially in FETT, where the $e_p$ index for the EU without the UK remained below 0.07 compared to 0.12-0.13 for the USA. This observation prompted us to investigate whether the low research level was common to all EU countries without the UK or there were great differences among them. For this purpose we calculated the $e_p$ index and P'$_{top\,0.01\%}$ indicator for the 16 most research-active EU countries without the UK (Table 5; this table does not include Poland for the reasons given in section 5). We calculated the fractional and domestic (all authors in the considered country or sets of countries) counts taking account of internal collaborations among EU countries without the UK but excluding external collaborations that would have increased their research performance. For reference, we included UK, Switzerland, and Singapore in this study, but computing only domestic publications in these countries, in order to compare them with domestic results for the other countries, otherwise the results would not be comparable.

Table 4. Research performance based on the $e_p$ index and P'$_{top\,0.01\%}$ indicator of four geographical research areas in slow evolving technology topics in three consecutive years[a]

| Research area | 2012 | | | 2013 | | | 2014 | | |
|---|---|---|---|---|---|---|---|---|---|
| | Papers | $e_p$ | P'$_{top\,0.01\%}$ | Papers | $e_p$ | P'$_{top\,0.01\%}$ | Papers | $e_p$ | P'$_{top\,0.01\%}$ |
| USA | 18573 | 0.128 | 5.03 | 19394 | 0.126 | 4.76 | 19344 | 0.115 | 3.40 |
| ERA | 41120 | 0.097 | 3.92 | 42556 | 0.091 | 3.30 | 41889 | 0.093 | 3.44 |
| Others | 77429 | 0.085 | 3.94 | 85847 | 0.089 | 5.22 | 91834 | 0.089 | 5.57 |
| EU w/o UK | 32414 | 0.095 | 2.76 | 33500 | 0.086 | 2.08 | 33242 | 0.089 | 2.34 |

[a] The P'$_{top\,0.01\%}$ indicator was calculated after fitting the percentile empirical data to a power law function. The $e_p$ index was calculated from de exponent of the power law equation, $e_p = 10^{-\alpha}$.

Attending first to the $e_p$ index, the most interesting finding was that only in The Netherlands was the $e_p$ index higher than 0.1— the world's $e_p$ index reference. In Ireland the $e_p$ index seemed to be intermediate between those of the UK and the top continental countries. Excluding the UK and Ireland, for the remaining countries, excluding Hungary, the $e_p$ index decreased from 0.08 to 0.05, approximately. Only in Austria and Ireland was the $e_p$ index clearly lower under fractional than under domestic counting. In the four biggest continental countries—Germany, France, Italy, and Spain—the $e_p$ index was below 0.08. Aside from minor differences that appear to be irrelevant for our study, the comparison of the $e_p$ index values of the four biggest EU countries (0.076-0.048) with those of Switzerland (0.151), and especially Singapore (0.196) confirms the poor performance of the EU's technological research. It is interesting to notice that small variations in $e_p$ reflect much higher differences in the likelihood of publishing very highly cited papers. Table 5 shows clearly this effect: The Netherlands and Sweden publish a similar number of papers, but have different $e_p$ index values, 0.118 and 0.072, respectively. Such a small change in $e_p$ produces a value of P'$_{top\ 0.01\%}$ which is 7.5 times larger in favor of The Netherlands.

Table 5. Research performance based on the $e_p$ index and P'$_{top\ 0.01\%}$ indicator of the most productive continental EU countries in fast evolving technological topics and a comparison with Switzerland, UK, and Singapore[a]

| Country | Number of papers | Fractional counts[b] | | Domestic counts | |
|---|---|---|---|---|---|
| | | $e_p$ index | P'$_{top\ 0.01\%}$ | $e_p$ index | P'$_{top\ 0.01\%}$ |
| Netherlands | 1499 | 0.118 | 0.250 | 0.111 | 0.141 |
| Ireland | 497 | 0.087 | 0.026 | 0.095 | 0.029 |
| Austria | 774 | 0.068 | 0.012 | 0.077 | 0.013 |
| Germany | 7480 | 0.078 | 0.259 | 0.076 | 0.13 |
| Finland | 781 | 0.077 | 0.022 | 0.076 | 0.015 |
| Denmark | 804 | 0.076 | 0.027 | 0.073 | 0.017 |
| Sweden | 1402 | 0.072 | 0.034 | 0.069 | 0.019 |
| Belgium | 1227 | 0.061 | 0.017 | 0.060 | 0.01 |
| Spain | 4061 | 0.060 | 0.051 | 0.057 | 0.032 |
| Portugal | 968 | 0.067 | 0.018 | 0.052 | 0.005 |
| Greece | 735 | 0.058 | 0.007 | 0.049 | 0.003 |
| Italy | 4320 | 0.051 | 0.037 | 0.048 | 0.024 |
| France | 5373 | 0.054 | 0.042 | 0.048 | 0.022 |
| Czech Republic | 909 | 0.047 | 0.003 | 0.046 | 0.002 |
| Hungary | 388 | 0.008 | 1.3E-06 | 0.007 | 5.6E-07 |
| | | | | | |
| Singapore | 3066 | – | – | 0.196 | 2.19 |
| Switzerland | 960 | – | – | 0.151 | 0.49 |
| UK | 3114 | – | – | 0.107 | 0.45 |

[a] Year 2014. The P'$_{top\ 0.01\%}$ indicator was calculated after fitting the percentile empirical data to a power law function. The $e_p$ index was calculated from de exponent of the power law equation, $e_p = 10^{-\alpha}$. Continental EU countries are sorted by their domestic $e_p$ index.
[b] Fractional counts: the search included only EU countries

Attending to the P'$_{top\ 0.01\%}$ indicator, the comparisons again show an overwhelmingly low competitiveness in the EU countries. Even considering the size of the system, the likelihood of publishing a domestic paper in the P$_{top\ 0.01\%}$ layer was 10 times higher for Singapore than for the four biggest EU countries combined, even though the total number of publications was ≈ 7 times lower in Singapore. This implies that, normalizing by the number of publications, Singapore is 70 times more efficient than the four biggest EU countries as a whole.

To show country differences independently of their size, Fig. 2 shows the theoretical number of papers in the P$_{top\ 0.01\%}$ layer if the total number of FETT papers were 10,000 in a selection of countries; Fig. 3 shows the cumulative probability function (equation [2] divided by *N*) for Germany and Singapore. Again, the inspection of these figures undoubtedly reveals the low level of competitiveness of EU research.

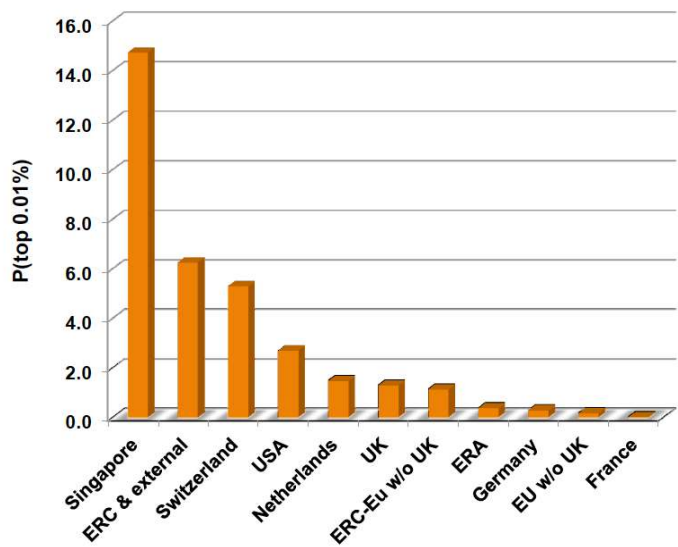

*Fig. 2. Graphical representation of the number of publications that would be in the top 0.01% most cited layer supposing a total production of 10,000 publications. Calculus with the values of the e$_p$ index in FETT in 2014 for selected countries and for European Research Council funded research. Abbreviations: ERA, European Research Area; ERC, funding by the European Research Council; ERC & external, funding by ERC but at least one collaborator does not belong to the ERA; EU w/o UK, EU countries excluding the UK. Values of the e$_p$ index for the cases shown in the figure, from left to right: 0.196, 0.158, 0.152, 0.128, 0.111, 0.107, 0.104, 0.080, 0.076, 0.067, 0.048.*

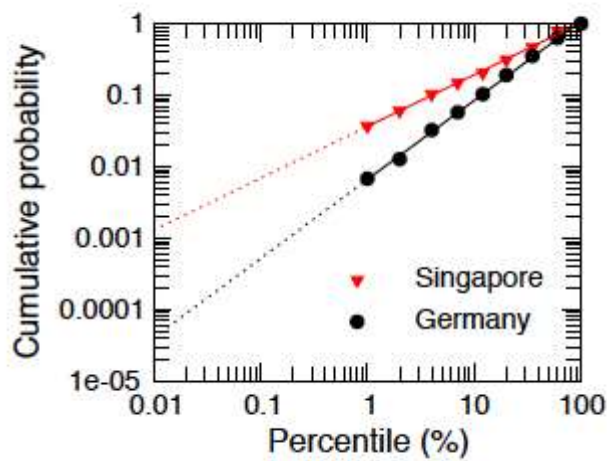

*Fig. 3. Plot of the cumulative probability function for Germany and Singapore. Data points for FETT in 2014. The fitted lines have been extended up to the 0.01% percentile to show the big differences between the two countries at this level. Values of the $e_p$ index: Singapore, 0.196; Germany, 0.076*

*6.2. The European Research Council funded research*

According to several documents from the European Commission, ERC funded projects are at the forefront of research excellence. For example: "As shown by the interim evaluation of Horizon 2020, the ERC has become a global beacon of scientific excellence" ((European-Commission, 2017b), p. 13).

Table 6. Research performance based on the $e_p$ index and P'$_{top\,0.01\%}$ indicator of publications in fast evolving technological topics acknowledging European Research Council funding support[a].

| Research area | Number of papers | $e_p$ index | P'$_{top\,0.01\%}$ |
|---|---|---|---|
| All | 2,146 | 0.139 | 1.42 |
| EU w/o UK | 841 | 0.104 | 0.20 |
| ERA | 1339 | 0.129 | 0.68 |
| ERA & external | 806 | 0.158 | 0.82 |

[a] Year 2014. EU w/o UK: authors from exclusively EU countries without the UK; ERA: authors from exclusively in ERA countries; ERA & external: publications including authors from any country, which implies that there is also at least one author from an ERA country.

In light of the results presented in Tables 3-5, this affirmation seemed doubtful, at least in technology and we tested it by again using the $e_p$ index and the P'$_{top\,0.01\%}$ indicator. The results (Table 6) show that the performance of ERC-funded research in FETT is better than the general performance of the EU (see Table 3 for comparison). However, the $e_p$ index of the publications from the EU without the UK is almost exactly 0.10, which invalidates any statement of excellence because 0.10 corresponds to the world average. Confirming once more the general conclusion of a low research performance across the EU, cooperation with other countries (at least one non-ERA country; ERA & external in Table 6) notably improved the efficiency of the ERC-funded research, raising the $e_p$ index to 1.58, which signifies an astonishing improvement. Although we did not study this issue, it seems that the most frequent collaborations were with USA, China, Japan, and Australia.

The remarkable increase in the $e_p$ index for ERC-funded research in non-ERA collaborations raises the question of whether it was due to the benefit of collaborating with countries with a superior research performance, or simply due to the increase of citations that might result from the wider audience for transcontinental publications. Although this issue has been addressed previously (Glänzel, 2001; Guerrero-Bote et al, 2013, Lancho-Barrantes, 2013 #1733) we performed a test in the landscape of our study. For this purpose we studied the case of Singapore, a very small country that we found to be at the top of research performance in FETT and which has a strong record of collaboration with China and the USA. Obviously, collaboration with these two much bigger countries should increase the research performance of Singapore, if the hypothesis of the wider audience was correct. The results summarized in Table 7 show that in Singapore international collaboration (whole or fractional counting) had a minimal effect on the $e_p$ index, but a notable influence on the P'$_{top\ 0.01\%}$ indicator as a consequence of the higher number of publications. Table 7 shows also the $R^2$ parameter of the fitting to demonstrate the statistical robustness of the comparisons. These results demonstrate that despite their wider audience and total number of citations, international publications does not increase the $e_p$ index of countries with a high *breakthrough potential*. They also demonstrate that even under the ERC's stringent project selection process the results of EU projects are much better when they are performed in collaboration with external countries whose research performance is higher; the hypothesis of the higher audience can be ruled out.

Table 7. Effects of international collaboration on the research performance indicators $e_p$ index and P'$_{top\ 0.01\%}$ in fast evolving technological topics in Singapore in 2014

| Counting procedure | Number of papers | $e_p$ index | P'$_{top\ 0.01\%}$ | $R^2$ parameter of fitting |
|---|---|---|---|---|
| Domestic | 1542 | 0.196 | 2.19 | 0.9987 |
| Fractional | 2450 | 0.197 | 3.58 | 0.9994 |
| Whole | 3066 | 0.198 | 5.27 | 0.9997 |

## 7. Discussion

*7.1. Technological research in the EU is far from excellent*

Our study aimed to demonstrate the weakness of research in the EU. For this purpose we used of the $e_p$ index, which measures the *intrinsic efficiency* or *breakthrough potential* of the research system and is a metric of excellence, and the P'$_{top\ 0.01\%}$ indicator to estimate the number of breakthroughs (explained in section 3). Moreover, our focus on technological research is justified because many claims about the scientific excellence of EU research are focused on technological topics that support the knowledge-based economy (Sorensen et al, 2016). The following are two such claims which explain our focus: "One of Europe's major weaknesses lies in its inferiority in terms of transforming the results of technological research and skills into innovations and competitive advantages" ((European-Commission, 1995), p. 5) and "The objective of Horizon 2020 is to reinforce and extend the excellence of the Union's science base and to consolidate the European Research Area in order to make the Union's research and innovation system more competitive on a global scale" ((European-Commission, 2017a), p. 106).

Future studies applying the same approach might reveal different findings in other research fields such as, for example, biomedicine, astrophysics, or particle physics.

Taking as a reference point a value of the $e_p$ index of 0.1, which is the world reference, the values of the $e_p$ index in the EU without the UK in FETT, around 0.07 (Table 3), reveal the weakness of the EU research and demonstrate the absence of research excellence in the forefront of technology. Examining the three geographical areas into which we divided the world, the USA, the EU, and Others, the values of the $e_p$ index show that the worst research performance is in the EU and the best is in the USA. The P'$_{top\ 0.01\%}$ indicator shows similar values for the USA and Others, which was due the much higher number of publications from Others. This indicates higher effort and investments to compensate for the lower *breakthrough potential*. In any case, this competition with the USA and Others leaves the EU in a third position of low relevance in relation to the number of possible technological breakthroughs.

Also disappointing are the $e_p$ index values in continental EU countries, which with the exception of The Netherlands, are all below 0.08 (Table 5). These data show that in Switzerland and especially in Singapore the *breakthrough potential* or excellence of the research performance in FETT is high and that in the continental EU countries there is no excellence. Regrettably, the P'$_{top\ 0.01\%}$ indicator reveals that the four biggest continental EU countries, which as a whole exceed 250 millions inhabitants, are far from being able to compete with Singapore which has less than 6 millions inhabitants. Considering the great difference in the number of papers, seven times higher in the four EU countries than in Singapore, a reasonable deduction is that technological research is profitable in Singapore and unprofitable in the EU.

In SETT the EU research performance is better than in FETT, but here too the values of the $e_p$ index are lower than 0.1, which implies that research is weak rather than excellent. Again, in SETT research in the USA shows the highest degree of excellence, but due to its lower number of publications relative to Others or the EU, the US's $P'_{top\ 0.01\%}$ indicator shows no large differences from the two other geographical blocks. Therefore, the role of the EU in global research is more relevant in SETT than in FETT.

The assumption that "the ERC has become a global beacon of scientific excellence" ((European-Commission, 2017b), p. 13) requires some qualifications when referring to hot technological topics. According to the results summarized in Table 6, the $e_p$ index of the EU without the UK, 0.104, implies no excellence. In contrast, considering all ERC publications or those from the whole ERA, the values of the $e_p$ index, 0.139 and 0.129, respectively, imply a certain level of excellence. However, the most interesting finding, which again questions the excellence of EU research, is that the $e_p$ index of ERC publications reaches its maximum value of 0.158 for collaborative publications in which there is at least one non-ERA country involved. This implies that ERC-funded publications are excellent when there is at least an author of an external country that participates and perhaps focuses the research. The hypothesis that the enhancing effect of an external country on the $e_p$ index is merely due to an increase of the citing researchers can be ruled out because in the same research field an increasing effect does not occur for Singapore (Table 7). If such an effect existed, it would be much higher in Singapore than in the EU because the population of internal researchers is much higher in the EU than in Singapore.

In summary, our results, which use an indicator that has a mathematical definition and that is based on the existing correlation between highly cited papers and important breakthroughs (Brito & Rodríguez-Navarro, 2018), corroborates previous studies (Albarrán et al, 2010; Bonaccorsi, 2007; Bonaccorsi et al, 2017a; Dosi et al, 2006; Herranz & Ruiz-Castillo, 2013; Rodriguez-Navarro & Narin, 2017; Sachwald, 2015) and demonstrates that EU research on hot technological topics is uncompetitive. Fig. 2 shows graphically what Ruiz-Castillo called "a truly European drama" (Ruiz-Castillo, 2016).

*7.2. Causes of the poor performance of EU research*

The status of technological research in the EU that we have described is overwhelmingly dire because not so long ago Europe was a beacon of progress. At the end of the 19th century Europe was an indisputable scientific and technological leader: "There is a dynamism about nineteenth century Europe that far exceeds anything previously known. Europe vibrated with power as never before: with technical power, economic power, cultural power. Its prime symbols were engines⎯the locomotives, the gasworks, the electrical dynamos" (Davies, 1997), p. 759). Starting from this position, the low efficiency of Germany and France at the forefront of technological research is particularly surprising.

Although the causes of this scientific decline deserve specific studies, it is hard to believe that something other than research policy is the cause of this decline. More than 10 years ago, two academic studies called attention to this decline: "the European picture shows worrying signs of weakness with respect to the generation of both scientific knowledge and technological innovation" (Dosi et al, 2006), p. 1461) and "This paper offers detailed evidence of the weak performance of European science in the upper tail of scientific quality, in fast moving scientific fields" (Bonaccorsi, 2007), p. 303). Despite these clear warnings the European Commission continued with its "wrong diagnoses and misguided policies" (Dosi et al, 2006), p. 1461).

Before trying to find an explanation for the weakness of technological research in the EU, it is worth noting that in slow evolving scientific fields, research quality in the EU and the USA is similar. This similarity has been demonstrated by double rank analysis in "plant sciences" (Rodríguez-Navarro & Brito, 2018), a field which might be an illuminating case study. A notable characteristic of "plant sciences" is that, although agriculture is essential for providing food to the inhabitants of our planet, the knowledge progress in plant sciences that supports it is insignificant in comparison with electronic technologies. This observation raises the question of whether there is an excessive focus of EU research in slow evolving fields, where EU research is more competitive. For example, it is surprising that the "mechanism of plant innate immunity" is a "hot research front" in the EU (European-Commission, 2017a), p. 57).

Considering the complex mechanisms through which researchers bring about scientific progress (Azoulay et al, 2011; Charlton, 2008; Jia et al, 2017) and the diversity of research policies in EU countries, the causes of the weak research performance in the EU might be diverse, though all having the same effect. Perhaps European researchers are less willing than others to take the risk of tackling the type of research that leads to important discoveries (Charlton, 2008).

In addition to this risk avoidance, the presumption of EU research excellence by the European Commission might have also promoted weak performance at the leading edge of knowledge in some countries. In the first place, this is because some national governments have internalized this assumption, reducing their interest in boosting research—why improve what is already excellent? In the second place because amid the euphoria about excellence the European Commission has failed to develop its member countries' research surveillance. The case of Spain provides an example: its traditionally low investments in academic research suffered a 50% reduction due to the economic crisis, which has produced a very significant dismantling of the research system that had been developed over more than 30 years. The reconstruction of the system will cost far more than the money saved and will take many years. Incomprehensibly, although Spain is a recipient of cohesion funds, this research policy was never censored by the European Commission.

An unlucky explanation for our results is that the distribution of the *Q* index, which captures the ability of scientists to succeed, (Sinatra et al, 2016) is worse in the researcher population of EU countries than in other countries. If this problem does not have a genetic basis, the selection of researchers in the EU was made improperly for many years, which again suggest a flawed research policy.

*7.3. Innovation and the economy are at risk in the EU*

As described above in sections 1 and 2, the central hypothesis of the EU research programs is the contrast between a low capacity in innovation and its excellent research: "The EU's innovation deficit is not due to a lack of knowledge or ideas, but because we do not capitalise on them. We need rapid European or international scale-up of innovative solutions." (European-Commission, 2017b), p. 11). Yet, in contrast with this optimistic view of excellent research, the present and previous studies show that technological research is not as excellent as the European Commission claims. This situation raises the question of whether the scenario is just the opposite: innovation is limited by the poor performance of technological research, which would make weak research the greatest problem for the future of the EU economy.

A high number of empirical studies have demonstrated that innovation entails much more than research but also that innovation depends on novel knowledge (e.g., (Etzkowitz & Leydesdorff, 2000; Leydesdorff & Meyer, 2006; Weitzman, 1998). On the other hand, the economic profitability of research has been demonstrated (Mansfield, 1990; Mansfield, 1998). The absence of a simple model for the economic benefits of research (Salter & Martin, 2001) can be explained by the great diversity in the efficiency of research across countries, which so far has not been taken into account. Our data show that the economic benefits of research cannot be the same in the EU, the USA, and Singapore.

The question which then arises is whether the EU can compete with other countries that produce much more efficient research. Although the answer to that question is not within the scope of this study, it is highly probable that the poor performance of the EU in technological research has had an important influence on innovation and competitiveness. Taking into consideration the low values of the $e_p$ index and $P'_{top\ 0.01\ \%}$ indicator for the EU's technological research, we can conclude that the innovation capacity of the EU's industry might not be low, and that it is rather the low technological research performance which limits innovation. From this point of view, the creation of the European Innovation Council seems to be based again on the wrong diagnosis that assumes that research institutions are producing breakthrough ideas but "often lack the in-house capabilities to nurture break-through innovations and their spinouts find it difficult to scaleup" (European-Commission, 2018), p. 14). It is worth insisting that the production of breakthrough ideas is much less frequent than in competitor countries.

Even if the reasons for the EU's record of weak innovation were in doubt, it should not perform the experiment of maintaining the current low research efficiency expecting to find out that innovation can be improved by other means. It is obvious that some non-EU countries are not going to stop improving their own research systems while they wait for the end of the EU experiment. Research activity in Singapore was insignificant 25 years ago, but over this period its output of papers has increased 25 times and, much more importantly, it now seems to have the highest level in the world on the $e_p$ index of FETT research. Singapore cannot be a threat to the EU economy because it is a small country with a population of less than 6 million, but the population of China exceeds 1,300 million, and China will not be waiting until the EU wakes up from its dream of excellent research.

It follows from these considerations that the EU should increase the efficiency of its research. Thus, the statement: "It is essential – also as a strong signal to the rest of the world – that both the EU and its Member States finally undertake to reach the 3 % target of GDP invested in R&I" (European-Commission, 2017b), p. 10) should be modified: "It is essential that both the EU and its States become the model of research efficiency." This does not mean that investments should not be increased, because although efficiency is not determined by investment it is not independent of it. Undoubtedly, the aforementioned reduction in research investments in Spain (section 7.2) will reduce research efficiency more than the number of its publications.

Making the analysis of research profitability more complex, effectiveness should also be considered. This issue can be summarized in the following question: Does the EU investigate what it needs? As discussed above (section 7.2), the answer to this question is probably no, because a large proportion of EU research projects is proposed by researchers themselves. There is no doubt that curiosity-driven research has let to important discoveries (Flexner, 2017) but this applies to specific cases not to all research production. In fact, in many cases much research that is alleged to be curiosity driven is in reality proposed under a high "publish or perish" pressure and merely produces sound papers (Rodríguez-Navarro, 2009). Furthermore, curiosity-driven research could address an almost infinite number of topics but the funds that society can allocate to research are finite (Weinberg, 1962). Obviously, we do not propose that research has to be programed by a state body. We only notice that not all research is equally beneficial for society.

Because most, if not all, of its problems with research seem to derive from "wrong diagnoses and misguided policies" as detailed more than 10 years ago (Dosi et al, 2006), p. 1461), the EU must accept the reality of its currently weak research system and restructure its research policy, focusing it on the correction of its failings. The revised policy should include the supervision of investments and research performance of member States. Otherwise, the economic future of Europeans is at risk.

**Acknowledgements**

We thank an anonymous reviewer for suggesting the term *breakthrough potential*, which we have adopted. This work was supported by the Spanish Ministerio de Economía y Competitividad, grant numbers FIS2014-52486-R and FIS2017-83709-R. This work was supported by the Spanish Ministerio de Economía y Competitividad, grant numbers FIS2014-52486-R and FIS2017-83709-R.